# Molecular dynamics with time dependent quantum Monte Carlo

Ivan P. Christov

Physics Department, Sofia University, 1164 Sofia, Bulgaria

## Abstract

In this paper we propose an *ab initio* method to solve quantum many-body problems of molecular dynamics where both the electronic and the nuclear degrees are represented by ensembles of trajectories and guiding waves in physical space. Both electrons and nuclei can be treated quantum mechanically where the guiding waves obey a set of coupled Schrodinger equations (quantum-quantum description) or, alternatively, coupled Schrödinger-Newtonian equations are solved for the quantum-classical approximation. The method takes into account local and non-local quantum correlation effects in a self consistent manner. The general formalism is applied to one- and two-dimensional hydrogen molecule subjected to a strong ultashort optical pulse. Comparison is made with the results from the "exact" Ehrenfest molecular dynamics for the molecular ionization and for the evolution of the internuclear distance as the molecule dissociates.

## 1. Introduction

The recent advent of femtosecond and attosecond experimental techniques has offered new possibilities to visualizing the electron motion inside atomic, molecular and condensed phase structures [1,2]. These techniques hold the potential to capture instants during chemical reaction dynamics where the different constituents are in complex transient states before the final state is established. It is even more intriguing that the outcome of some chemical reactions could be controlled in predictable way using appropriately synthesized ultrashort optical pulses. These opportunities stimulated significant theoretical efforts directed towards developing advanced methods to describe electron and nuclear dynamics in complex multielectron systems [3,4], that was also promoted by the rapid progress in the computational capabilities. Unfortunately the direct numerical solution of the many body time-dependent Schrödinger equation remains feasible only for simple few-body quantum systems. Therefore a variety of approximate methods has been adopted for performing practical calculations. These include time-dependent density functional approximation [5,6], multi-configuration time-dependent Hartree-Fock theory [7,8], time-dependent configuration interaction [9,10], and many-body Green function [11] techniques. While describing well the correlated few-electron motion within atoms, these many-body quantum methods usually ignore the nuclear dynamics and assume a fixed internuclear distance in molecules [12-14]. Other approaches introduce certain ansatz to couple the electronic and nuclear degrees of freedom with the aim to compute the dynamics of both electrons and nuclei, simultaneously, e.g. [15-17]. Together with these techniques, which are based mostly on evolution of wavefunctions, there exist also particle-based and mixed quantum-classical methods where the heavy particles are treated classically. The coupling between the classical and the quantum degrees has been addressed by mean-field and trajectory-surface hopping approaches [18].

The molecular dynamics (MD) approach is computationally advantageous and has proven extremely valuable for calculating the dynamics of elementary chemical processes, e.g. [19]. It has been used to calculate the response of simple molecules to intense ultrashort laser fields where the influence of the nuclear motion on the produced high-harmonic spectrum is of importance [20]. In these “beyond Born-Oppenheimer” models the classical particles (nuclei) influence the quantum subsystem through explicit coordinate dependence in the potential term of the time-dependent Schrödinger equation (TDSE), which constitutes the forward reaction problem. At the same time the classical particles experience a mean backward force from the quantum subsystem through the gradient of the average potential. This approach to the back-reaction problem is known as Ehrenfest molecular dynamics, e.g. [21], where however the detailed quantum correlations between the light and the heavy particles are neglected. It is clear that there is certain asymmetry in the way the forward and backward reaction problems are treated in this approximation, which is based solely on the different masses of the electrons and the nuclei. One different approach to the back-reaction problem in quantum-classical approximation is to take two ensembles of particles where each member of the electron ensemble is described through a Bohmian trajectory which is coupled to a single heavy particle. Then, the sets of quantum and classical particles are propagated in time using the Schrödinger and the Newtonian equations, respectively [22-23]. Here we consider a new “quantum-quantum” description of the dynamics of systems of light and heavy particles, where specific quantum effects of electron-nuclear and nuclear-nuclear correlations can be captured. Our approach is based on the recently proposed time dependent quantum Monte Carlo method where each microscopic physical particle is described by an ensemble of fictitious classical particles (walkers) and a corresponding ensemble of quantum waves, where the particles and the waves participate in the dynamics on an equal footing[24-26]. Mathematically the TDQMC method is expressed as a set of coupled time dependent Schrödinger equations for the guiding waves in physical space while the walkers move in the same space according to the de Broglie-Bohm guiding equation, with no quantum potentials involved. In this formulation the density of walkers reproduces at each instant the density of physical particles (electrons or nuclei) while the guiding waves preserve their statistical interpretation of quantum mechanics in physical space. In this paper we use the TDQMC methodology to describe the time evolution of a system of electrons and nuclei which interact through potentials, where we use the "effective potential" concept to describe the nonlocal quantum correlation effects in the system. We apply the method to simulate the ionization and dissociation of one-dimensional hydrogen molecule subjected to a strong optical pulse, and the orientation of 2D hydrogen molecule.

## 2. Deriving TDQMC molecular dynamics

For a non-relativistic system consisting of K nuclei and N electrons the Schrödinger equation reads:

$$i\hbar\frac{\partial}{\partial t}\Psi(\mathbf{R},\mathbf{r},t)=H\Psi(\mathbf{R},\mathbf{r},t)\,, \tag{1}$$

where:

$$H = -\sum_{I}^{K} \frac{\hbar^2}{2M_I} \nabla_I^2 - \sum_{i}^{N} \frac{\hbar^2}{2m_e} \nabla_i^2 + V\left(\mathbf{R}, \mathbf{r}, t\right), \qquad (2)$$

is the many-body quantum Hamiltonian, and $\mathbf{R} = \{\mathbf{R}_1, ..., \mathbf{R}_K\}$ and $\mathbf{r} = \{\mathbf{r}_1, ..., \mathbf{r}_N\}$ are the nuclear and the electronic degrees of freedom, respectively. The Hamiltonian in Eq. (2) is a sum of electron-nuclear, nuclear-electron, electron-electron, nuclear-nuclear, and external potentials:

$$\begin{aligned} V(\mathbf{R}_1, ..., \mathbf{R}_K, \mathbf{r}_1, ..., \mathbf{r}_N, t) &= V_{e-n}(\mathbf{R}_1, ..., \mathbf{R}_K, \mathbf{r}_1, ..., \mathbf{r}_N) + V_{e-e}(\mathbf{r}_1, ..., \mathbf{r}_N) + V_{n-n}(\mathbf{R}_1, ..., \mathbf{R}_K) \\ &\quad + V_{ext}(\mathbf{R}_1, ..., \mathbf{R}_K, \mathbf{r}_1, ..., \mathbf{r}_N, t) \\ &= \sum_{i \geq J} V_{e-n}(\mathbf{r}_i - \mathbf{R}_J) + \sum_{I > j} V_{n-e}(\mathbf{R}_I - \mathbf{r}_j) + \sum_{i > j} V_{e-e}(\mathbf{r}_i - \mathbf{r}_j) + \sum_{I > J} V_{n-n}(\mathbf{R}_I - \mathbf{R}_J) \\ &\quad + V_{ext}(\mathbf{R}_1, ..., \mathbf{R}_K, t) + V_{ext}(\mathbf{r}_1, ..., \mathbf{r}_N, t) \end{aligned} \qquad (3)$$

where the electron-nuclear $V_{e-n}$ and nuclear-electron $V_{n-e}$ potentials have identical coordinate dependence and these are separated in Eq. (3) for convenience. Simultaneous self-consistent evolution of the electronic and nuclear degrees is necessary in order to account for the nonadiabatic changes in the quantum states due to possible non-Born-Oppenheimer dynamics in ultrafast laser fields. To this end we start with a single configuration ansatz (single Hartree product) for the total wave function where the nuclear and electronic coordinates are separated $\Psi(\mathbf{R}, \mathbf{r}, t) = \Phi(\mathbf{R}, t) \cdot \varphi(\mathbf{r}, t)$ [18], where in addition we assume full factorization of the many-body wavefunctions for the nuclei $\Phi(\mathbf{R},t)$ and for the electrons $\varphi(\mathbf{r},t)$ (see also [27, 28]):

$$\Psi(\mathbf{R},\mathbf{r},t)=\prod_{I=1}^{K}\Phi_I(\mathbf{R}_I,t)\cdot\prod_{i=1}^{N}\varphi_i(\mathbf{r}_i,t), \tag{4}$$

In Eq. (4) it is assumed that the separate nuclear and electronic wave functions are normalized to unity at every instant of time. Following the standard procedure of Hartree theory we then arrive at a set of coupled non-linear integro-differential equations for the nuclear and electronic wave functions:

$$i\hbar\frac{\partial}{\partial t}\Phi_I(\mathbf{R}_I,t)=\left[-\frac{\hbar^2}{2M_I}\nabla_I^2+\sum_{J\neq I}\int d\mathbf{R}_J V_{n-n}(\mathbf{R}_I-\mathbf{R}_J)\left|\Phi_J(\mathbf{R}_J,t)\right|^2\right.$$
$$\left.+\sum_j\int d\mathbf{r}_j V_{n-e}(\mathbf{R}_I-\mathbf{r}_j)\left|\varphi_j(\mathbf{r}_j,t)\right|^2+V_{ext}(\mathbf{R}_I,t)\right]\Phi_I(\mathbf{R}_I,t) \tag{5}$$

$$i\hbar\frac{\partial}{\partial t}\varphi_i(\mathbf{r}_i,t)=\left[-\frac{\hbar^2}{2m_e}\nabla_i^2+\sum_{j\neq i}\int d\mathbf{r}_j V_{e-e}(\mathbf{r}_i-\mathbf{r}_j)\left|\varphi_j(\mathbf{r}_j,t)\right|^2\right.$$
$$\left.+\sum_J\int d\mathbf{R}_J V_{e-n}(\mathbf{r}_i-\mathbf{R}_J)\left|\Phi_J(\mathbf{R}_J,t)\right|^2+V_{ext}(\mathbf{r}_i,t)\right]\varphi_i(\mathbf{r}_i,t), \tag{6}$$

where in Eq. (5) and Eq. (6) $I=1,\dots,K$, $i=1,\dots,N$, and we have omitted terms which do not depend on $\mathbf{R}_I$ in Eq. (5) and on $\mathbf{r}_i$ in Eq. (6), respectively, because these terms do not influence the motion of the corresponding Monte Carlo walkers (see Eq. (8) and Eq. (9)

below). Note that Eq. (5) and Eq. (6) differ from the standard self-consistent field equations [18] in that here we use one-body wavefunctions.

It is known that the single determinant ansatz (Hartree) approximation and the resulting mean-field equations (Eq. (5) and Eq. (6)) disregard important local and nonlocal quantum correlation effects. One approach to overcome these difficulties is to use multi-configuration ansatz where multiple wavefunctions for the electron degree are used, which leads to multi-configuration time-dependent self-consistent field theory [29]. Here we employ the TDQMC methodology which assigns a separate set of wavefunctions and walkers to each physical particle where the wavefunctions guide the Monte-Carlo walkers, for both the electron and the nuclear degrees. Although we use a single configuration ansatz of Eq. (4), the large set of replicas of the quantum system generated for the different positions of the walkers and the guiding waves allows us to calculate local and non-local quantum correlation effects without invoking explicit series expansions over multiple configurations (that would involve time-expensive calculation of various integrals). This is especially important for large deformations of the electron cloud and/or for strong rotation and translation of the molecule (e.g. in an external field), where the standard multi-configuration expansions may converge extremely slow. As the system evolves towards stationary state the ensembles of waves and walkers tend to equilibrium via fluctuations, where at equilibrium the particle distribution functions obey $P_e(\mathbf{r},t)=|\varphi(\mathbf{r},t)|^2$ for the electrons and $P_n(\mathbf{R},t)=|\Phi(\mathbf{R},t)|^2$ for the nuclei [30,31]. At the same time, the local and non-local quantum correlation effects are incorporated into the equations for the walkers and for the guiding waves. For example, the exchange interaction between identical particles can be readily accounted for within TDQMC by introducing the symmetry properties of the many-body wavefunction in the guiding equations for the walkers. The many-body quantum state of the electrons can then be represented as an antisymmetrized product (single Slater determinant or a sum of Slater determinants):

$$\varphi(\mathbf{r},t) = A\prod_{i=1}^{N}\varphi_i(\mathbf{r}_i,t)\,, \tag{7}$$

which is next substituted into the de Broglie-Bohm guidance equation for the velocity field of the electron walkers [32]:

$$\mathbf{v}(\mathbf{r}_i^k) = \frac{\hbar}{m_e}\mathrm{Im}\left[\frac{1}{\varphi(\mathbf{r},t)}\nabla_i\varphi(\mathbf{r},t)\right]_{\mathbf{r}_j=\mathbf{r}_j^k(t)} \tag{8}$$

where $\mathbf{r}_j^k(t)$ is the trajectory of the k-th walker from j-th electron ensemble. Similarly, the equation for the velocity field of the nuclei can be written as:

$$\mathbf{V}(\mathbf{R}_I^k) = \frac{\hbar}{M_I}\mathrm{Im}\left[\frac{1}{\Phi(\mathbf{R},t)}\nabla_I\Phi(\mathbf{R},t)\right]_{\mathbf{R}_J=\mathbf{R}_J^k(t)} \tag{9}$$

where $\mathbf{R}_J^k(t)$ is the trajectory of the k-th walker from J-th nuclear ensemble. In addition, the two ensembles of walkers experience a random drift that thermalizes their distributions while complying with the nodes of the wavefunction via importance sampling [25]. For the electrons, the walker's coordinate is updated according to:

$$d\mathbf{r}_i^k = \mathbf{v}(\mathbf{r}_i^k)dt + \boldsymbol{\eta}_e\sqrt{\frac{\hbar}{m_e}dt} \quad , \tag{10}$$

where $\boldsymbol{\eta}_e$ is a vector random variable with zero mean whose variance decreases as we approach the ground state of the system. Similar equation holds for the updated coordinates in the nuclear ensemble:

$$d\mathbf{R}_I^k = \mathbf{V}(\mathbf{R}_I^k)dt + \boldsymbol{\eta}_n\sqrt{\frac{\hbar}{M_I}dt} \tag{11}$$

Another important quantum correlation effect which is neglected in Hartree approximation is related to a specific quantum nonlocality that arises due to dependence of many-body wavefunction in Eq. (1) on the coordinates in 3(K+N) dimensional configuration space. This nonlocality is evidenced as an interaction between different points in configuration space which represent the momentary coordinates of different replicas of the quantum system. One simple and efficient way to account for these quantum effects is to formally represent the particle densities in Eq. (5) and Eq. (6) by smoothed interpolation with e.g. Gaussian kernels that are centered at the positions of the Bohmian walkers (kernel density estimation) [26]. For the electrons we have:

$$\left|\varphi_j(\mathbf{r}_j,t)\right|^2 = \sum_{k=1}^{M}\frac{1}{z_j^k}\exp\left(-\frac{\left|\mathbf{r}_j - \mathbf{r}_j^k(t)\right|^2}{\sigma_j^k\left(\mathbf{r}_j^k,t\right)^2}\right), \tag{12}$$

and for the nuclei:

$$\left|\Phi_J(\mathbf{R}_J,t)\right|^2 = \sum_{k=1}^{M}\frac{1}{Z_J^k}\exp\left(-\frac{\left|\mathbf{R}_J - \mathbf{R}_J^k(t)\right|^2}{\Sigma_J^k\left(\mathbf{R}_J^k,t\right)^2}\right), \tag{13}$$

where M is the number of Bohmian walkers; $z_j$ and $Z_J$ are weighting factors to preserve the norm of the states for the electrons and the nuclei, respectively. Substituting Eq. (12) and Eq. (13) into Eq. (5) and Eq. (6), and assigning a separate guiding wave to each Bohmian walker, we transform the non-linear Hartree equations (5) and (6) to a set of coupled linear Schrödinger equations for the guiding waves:

$$i\hbar\frac{\partial}{\partial t}\Phi_I^k(\mathbf{R}_I,t) = \left[-\frac{\hbar^2}{2M_I}\nabla_I^2 + \sum_{j=1}^{N}V_{n-e}^{eff}[\mathbf{R}_I - \mathbf{r}_j^k(t)] \right.$$

$$\left. + \sum_{J\neq I}^{K}V_{n-n}^{eff}[\mathbf{R}_I - \mathbf{R}_J^k(t)] + V_{ext}(\mathbf{R}_I,t)\right]\Phi_I^k(\mathbf{R}_I,t) \tag{14}$$

$$i\hbar\frac{\partial}{\partial t}\varphi_i^k(\mathbf{r}_i,t) = \left[-\frac{\hbar^2}{2m_e}\nabla_i^2 + \sum_{J=1}^{K}V_{e-n}^{eff}[\mathbf{r}_i - \mathbf{R}_J^k(t)] \right.$$

$$\left. + \sum_{j\neq i}^{N}V_{e-e}^{eff}[\mathbf{r}_i - \mathbf{r}_j^k(t)] + V_{ext}(\mathbf{r}_i,t)\right]\varphi_i^k(\mathbf{r}_i,t), \tag{15}$$

where the nonlocal effective potentials are calculated as sums over the smoothed walker distributions:

$$V_{e-e}^{eff}[\mathbf{r}_i - \mathbf{r}_j^k(t)] = \frac{1}{z_j^k}\sum_{l=1}^{M} V_{e-e}[\mathbf{r}_i - \mathbf{r}_j^l(t)]\exp\left(-\frac{\left|\mathbf{r}_j^l(t) - \mathbf{r}_j^k(t)\right|^2}{\sigma_j^k\left(\mathbf{r}_j^k, t\right)^2}\right), \tag{16}$$

$$V_{e-n}^{eff}[\mathbf{r}_i - \mathbf{R}_J^k(t)] = \frac{1}{Z_J^k}\sum_{l=1}^{M} V_{e-n}[\mathbf{r}_i - \mathbf{R}_J^l(t)]\exp\left(-\frac{\left|\mathbf{R}_J^l(t) - \mathbf{R}_J^k(t)\right|^2}{\Sigma_J^k\left(\mathbf{R}_J^k, t\right)^2}\right), \tag{17}$$

$$V_{n-e}^{eff}[\mathbf{R}_I - \mathbf{r}_j^k(t)] = \frac{1}{z_j^k}\sum_{l=1}^{M} V_{n-e}[\mathbf{R}_I - \mathbf{r}_j^l(t)]\exp\left(-\frac{\left|\mathbf{r}_j^l(t) - \mathbf{r}_j^k(t)\right|^2}{\sigma_j^k\left(\mathbf{r}_j^k, t\right)^2}\right), \tag{18}$$

$$V_{n-n}^{eff}[\mathbf{R}_I - \mathbf{R}_J^k(t)] = \frac{1}{Z_J^k}\sum_{l=1}^{M} V_{n-n}[\mathbf{R}_I - \mathbf{R}_J^l(t)]\exp\left(-\frac{\left|\mathbf{R}_J^l(t) - \mathbf{R}_J^k(t)\right|^2}{\Sigma_J^k\left(\mathbf{R}_J^k, t\right)^2}\right), \tag{19}$$

where:

$$z_j^k = \sum_{l=1}^{M}\exp\left(-\frac{\left|\mathbf{r}_j^l(t) - \mathbf{r}_j^k(t)\right|^2}{\sigma_j^k\left(\mathbf{r}_j^k, t\right)^2}\right), \tag{20}$$

$$Z_J^k = \sum_{l=1}^{M}\exp\left(-\frac{\left|\mathbf{R}_J^l(t) - \mathbf{R}_J^k(t)\right|^2}{\Sigma_J^k\left(\mathbf{R}_J^k, t\right)^2}\right) \tag{21}$$

are the weighting factors. In fact, the effective potentials in Eq. (16) and Eq. (17) describe the weighted nonlocal Coulomb interaction experienced by a given trajectory from the *i*-th electron ensemble from the trajectories that belong to the *j*-th electron ensemble and from those from to the *J*-th nuclear ensemble. The width of the Gaussian kernel $\sigma_j^k\left(\mathbf{r}_j^k,t\right)$ plays the role of characteristic length of the nonlocal quantum correlations that depends on the electron density (the density of walkers) in the quantum system. At space locations where the walker density is higher the electron correlation length $\sigma_j^k\left(\mathbf{r}_j^k,t\right)$ in Eq. (16) and Eq. (18) is smaller in order to compensate for the higher number of interpolating Gaussians at that location. In these regions there are more intense interactions between the k-the walker from j-th electron ensemble and the walkers that represent the rest of electrons and nuclei. Because of the symmetry between the equations for electrons and nuclei (Eq. (14) and Eq. (15)) similar considerations hold for the nuclear nonlocal correlation length $\Sigma_J^k\left(\mathbf{R}_J^k,t\right)$. The nonlocal correlation lengths $\sigma_j^k\left(\mathbf{r}_j^k,t\right)$ and $\Sigma_J^k\left(\mathbf{R}_J^k,t\right)$ are not free parameters and can be estimated using simple formulae [33,34] :

$$\sigma_j^k\left(\mathbf{r},t\right)=\sigma\sqrt{\frac{g_j}{\rho_j^k(\mathbf{r},t)}}\,, \tag{22}$$

$$\Sigma_J^k\left(\mathbf{R},t\right)=\Sigma\sqrt{\frac{G_J}{\mathrm{P}_J^k(\mathbf{R},t)}}\,, \tag{23}$$

where $\rho_j^k(\mathbf{r},t)$ and $\mathrm{P}_J^k(\mathbf{R},t)$ are pilot density estimates of the walker distributions for the *j*-th electron and the *J*-th nucleus, which can be obtained using kernel density estimation with constant bandwidths σ and $\Sigma$, and $g_j$ and $G_J$ are the geometric means of the values of $\rho_j^k(\mathbf{r},t)$ and $\mathrm{P}_J^k(\mathbf{R},t)$, for $k=1,\ldots,M$, respectively. Also, in some representations $\sigma_j^k\left(\mathbf{r}_j^k,t\right)$ and $\Sigma_J^k\left(\mathbf{R}_J^k,t\right)$ are related to the covariance matrices for the corresponding walkers in two or three dimensional space [25, 34]. Note that kernel density estimations with Bohmian walkers has been used previously to find the parameters of a set of Gaussians that best approximate the probability density function [35,36], and for constructing and propagating in phase space quantum distribution functions such as the Wigner function in Husimi representation [37]. Equations (14) and (15) represent the quantum-quantum version of TDQMC molecular dynamics where both electronic and nuclear degrees are treated by sets of coupled Schrodinger equations, and by the corresponding guiding equations for the Monte Carlo walkers, Eq. (8) and Eq. (9).

## 3. Limiting cases

Different approximations to the quantum-quantum description can be derived from Eq. (14) and Eq. (15). If we disentangle the system replicas for the nuclear degree by letting $\Sigma_J^k = 0$ (ultra-correlated nuclei, see [25]), substitute the standard ansatz of Bohmian mechanics:

$$\Phi_I\left(\mathbf{R}_I,t\right)=R(\mathbf{R}_I,t)\exp[iS(\mathbf{R}_I,t)/\hbar] \tag{24}$$

into the wave equations for the nuclear degree, Eq. (14), and allow $\hbar\to 0$, we arrive at:

$$M_I\ddot{\mathbf{R}}_I^k=-\nabla_I\left[\sum_{I\neq J}V_{n-n}\left[\mathbf{R}_I-\mathbf{R}_J^k(t)\right]+\sum_j V_{n-e}^{eff}\left[\mathbf{R}_I-\mathbf{r}_j^k(t)\right]+V_{ext}(\mathbf{R}_I,t)\right], \tag{25}$$

which is the Newtonian equation of motion for a regular bundle of classical nuclear trajectories, where there is no quantum diffusion and the back-reaction is described by the effective nuclear-electron potential. The approximation given by Eq. (15) and Eq. (25) can be labeled "mixed quantum-classical TDQMC molecular dynamics". It corresponds to zeroed quantum potential for the nuclear degree, where the transition $\hbar\to 0$ has to be interpreted with some caution (see [32], Ch.6 for detailed discussion). Provided, in addition, that ultra-correlated electrons are assumed ($\sigma_j^k=0$), Eq. (25) is transformed to the case of Ref. 23. If we do not let $\hbar\to 0$ in the derivation of Eq. (25) the mixed quantum-classical Bohmian method of Ref. 22 is recovered. However, unlike in [22] here we do not use quantum potentials whose calculation may pose serious problems, especially for regions with lower density of particles. Different semi-classical corrections to Eq. (25) are also possible using expansion in powers of $\hbar$ in Eq. (14) and making use of Eq. (24) (see [38]). The Ehrenfest molecular dynamics can be recovered from Eq. (14) by letting $\Sigma_J^k=0$ and $\sigma_j^k\to\infty$, which leads to disentangled nuclei and mean-field electron-electron $V_{e\text{-}e}$ and nuclear-electron $V_{n\text{-}e}$ potentials in Eq. (15) and Eq. (25). This reduces the number of active degrees of freedom where we have one nuclear walker for each nucleus and one electronic guiding wave for each electron in Eq. (15). The electronic walkers play an auxiliary role in Ehrenfest approximation where they can be used to ease the calculation of the mean-field integrals by reducing them to Monte Carlo summing. The approximations considered here are to be compared with the "exact" Ehrenfest molecular dynamics where the electron motion is treated without approximation by a direct numerical solution of the time-dependent Schrödinger equation in configuration space while the nuclei obey the Newtonian equations with mean-field electron-nuclear potential $V_{e\text{-}n}$, e.g. [20]. It should be noted, however, that the "exact" Ehrenfest molecular dynamics suffers severe dimensionality bottleneck that limits its practical importance to up to three electrons in one spatial dimension.

**4. Energy estimations**

An estimate for the energy of a conservative system of interacting electrons and nuclei can be obtained as an average over the electronic and nuclear ensembles of walkers which represent the different physical particles:

$$E = \frac{1}{M}\sum_{k}^{M} E_k \ , \tag{26}$$

where $E_k$ is the energy of the $k$-th replica of the many-body system:

$$E_k = \sum_i \frac{1}{2} m_e \dot{\mathbf{r}}_i^{k2} + \sum_I \frac{1}{2} M_I \dot{\mathbf{R}}_I^{k2} + \sum_{i,I} Q(\mathbf{r}_1,...,\mathbf{r}_i^k,...,\mathbf{r}_N,\mathbf{R}_1...,\mathbf{R}_I^k,...,\mathbf{R}_K,t)$$

$$+\sum_{i>j} V_{e-e}(\mathbf{r}_i^k - \mathbf{r}_j^k) + \sum_{I>J} V_{n-n}(\mathbf{R}_I^k - \mathbf{R}_J^k) + \sum_{i\geq J} V_{e-n}(\mathbf{r}_i^k - \mathbf{R}_J^k) + \sum_{I>j} V_{n-e}(\mathbf{R}_I^k - \mathbf{r}_j^k) \ , \qquad (27)$$

where the quantum potential has been estimated for the trajectories $\left(\mathbf{r}_i^k(t),\mathbf{R}_I^k(t)\right)$:

$$Q(\mathbf{r}_1,...,\mathbf{r}_i^k,...,\mathbf{r}_N,\mathbf{R}_1...,\mathbf{R}_I^k,...,\mathbf{R}_K,t) =$$

$$-\frac{\hbar^2}{2m_e}\left[\frac{\nabla_i^2|\Psi(\mathbf{R},\mathbf{r},t)|}{|\Psi(\mathbf{R},\mathbf{r},t)|}\right]_{\mathbf{r}_i=\mathbf{r}_i^k(t)} - \frac{\hbar^2}{2M_I}\left[\frac{\nabla_I^2|\Psi(\mathbf{R},\mathbf{r},t)|}{|\Psi(\mathbf{R},\mathbf{r},t)|}\right]_{\mathbf{R}_I=\mathbf{R}_I^k(t)} \qquad (28)$$

Adaptive kernel density estimation can be employed to estimate the many-body quantum potential in Eq. (28) without referencing to numerical derivatives of multi-variate functions [25]. Having calculated the density distributions for the electrons and for the nuclei, the energy of each replica of the quantum system in stationary state at instant τ (where $\dot{\mathbf{r}}_i^k(\tau) = \dot{\mathbf{R}}_I^k(\tau) = 0$) is given by:

$$E_k = \left[\frac{\hbar^2}{8m_e}\sum_i \frac{\left[\nabla_i P_e(\mathbf{r})\right]^2}{P_e(\mathbf{r})^2} + \frac{\hbar^2}{8}\sum_I \frac{\left[\nabla_I P_n(\mathbf{R})\right]^2}{M_I P_n(\mathbf{R})^2} + \sum_{i>j} V_{e-e}(\mathbf{r}_i^k - \mathbf{r}_j^k)\right.$$

$$+\sum_{I>J} V_{n-n}(\mathbf{R}_I^k - \mathbf{R}_J^k) + \sum_{i \geq J} V_{e-n}(\mathbf{r}_i^k - \mathbf{R}_J^k) + \sum_{I>j} V_{n-e}(\mathbf{R}_I^k - \mathbf{r}_j^k) \Bigg] _{\substack{\mathbf{r}_{i,j}^k = \mathbf{r}_{i,j}^k(\tau) \\ \mathbf{R}_{I,J}^k = \mathbf{R}_{I,J}^k(\tau)}} \tag{29}$$

A simplified but still accurate estimation for the energy can be derived if we ignore the irreducibility of the many-body quantum potential assuming factorization of the many-body wave functions, as done in Eq. (4):

$$E_k = \Bigg[ \frac{\hbar^2}{8m_e} \sum_i \frac{\left[\nabla_i P_{ei}^k(\mathbf{r}_i^k)\right]^2}{P_{ei}^k(\mathbf{r}_i^k)^2} + \frac{\hbar^2}{8} \sum_I \frac{\left[\nabla_I P_{nI}^k(\mathbf{R}_I^k)\right]^2}{\mathrm{M_I} P_{nI}^k(\mathbf{R}_I^k)^2} + \sum_{i>j} V_{e-e}(\mathbf{r}_i^k - \mathbf{r}_j^k)$$

$$+\sum_{I>J} V_{n-n}(\mathbf{R}_I^k - \mathbf{R}_J^k) + \sum_{i \geq J} V_{e-n}(\mathbf{r}_i^k - \mathbf{R}_J^k) + \sum_{I>j} V_{n-e}(\mathbf{R}_I^k - \mathbf{r}_j^k) \Bigg] _{\substack{\mathbf{r}_{i,j}^k = \mathbf{r}_{i,j}^k(\tau) \\ \mathbf{R}_{I,J}^k = \mathbf{R}_{I,J}^k(\tau)}} , \tag{30}$$

where $P_{ei}^k(\mathbf{r}_i^k) = \left|\varphi_i^k(\mathbf{r}_i^k)\right|^2$ and $P_{nI}^k(\mathbf{R}_I^k) = \left|\Phi_I^k(\mathbf{R}_I^k)\right|^2$.

**5. Numerical results**

For illustration purposes here we calculate the ground state and the evolution of one-dimensional hydrogen molecule under the influence of strong ultrashort optical pulse. The results are then compared with the predictions from the "exact" Ehrenfest molecular dynamics, and with the time dependent Hartree-Fock approximation. In all cases non-adiabaticity well beyond Born-Oppenheimer approximation is assumed where both the electrons and the nuclei move in a self-consistent manner. The enhanced electron collision rate in one spatial dimension makes this model to be especially suitable for studying the electron-electron correlation effects and their role for molecular ionization and dissociation. One can also estimate the effects of quantum non-locality on these processes by tracking the evolution of the correlation distances σ and Σ as the system evolves in time. Our calculations use modified Coulomb potentials to avoid numerical complications coming from the singularity at the position of the nucleus [39]. We assume that the electron-nuclear and electron-electron interactions are approximated by the following potentials:

$$V_{e-n}(x_i - X_I) = V_{n-e}(X_I - x_i) = -\frac{e^2}{\sqrt{a + (x_i - X_I)^2}}\,; \tag{31}$$

$$V_{e-e}(x_i - x_j) = \frac{e^2}{b + \left|x_i - x_j\right|}\,, \tag{32}$$

$$V_{n-n}(X_I - X_J) = \frac{e^2}{\sqrt{c + (X_I - X_J)^2}} \tag{33}$$

where $I,J,i,j$=1,2 and we have chosen $a$=1 a.u. (atomic units), $b$=1.5 a.u., and $c$=0.5 a.u. in Eqs. (31) through (33). For hydrogen molecule the two electrons are shared between the nuclei and the two-body wavefunction in Eq. (7) can be represented as a symmetrized product of the two one-electron guiding waves, for each couple of electronic walkers. We propagare the walkers and the guiding waves by numerically solving Eq. (14) and Eq. (15) (or Eq. (25)) using a real space grid method with spatial extend of 50 a.u. with time step size of 0.1 a.u. As an initial condition we assign a separate guiding wave $\varphi_i^k(x_i, t=0) = \exp\left(-(x_i - 0.5)/\sigma_e^2\right) + \exp\left(-(x_i + 0.5)/\sigma_e^2\right)$ to each electronic walker where $\sigma_e$=2 a.u., and a guiding wave $\Phi_I^k(X_I, t=0) = \exp\left(-(X_I \pm 0.5)/\Sigma_n^2\right)$ to each nuclear walker, where $\Sigma_n$=0.5 a.u.. The initial distribution of the electronic walkers is determined by a Meropolis procedure to sample $\left|\varphi_i^k(x_i, t=0)\right|^2$. The initial distribution of nuclear walkers depends on the approximation we use. For quantum-classical simulation (Eq. (15) and Eq. (25)) there is no guiding waves for the nuclear degree and the nuclear walkers are positioned at -0.5 a.u. and +0.5 a.u., with no initial dispersion. For the quantum-quantum simulation (Eq. (14) and Eq. (15)) the nuclear walkers are distributed in space using a Metropolis procedure to sample $\left|\Phi_I^k(X_I, t=0)\right|^2$.

In order to obtain the walker distribution for the ground state and the energy of the molecule we propagate the electronic and the nuclear guiding waves over 2000 complex time steps in Eq. (14) and Eq. (15), together with evolving in real time the corresponding 2x200 Bohmian walkers through Eq. (8) and Eq. (9). The complex time ensures a nonzero velocity of the walkers in Eqs. (8)-(9) for evolution towards the ground state [25]. Other methods to calculate the ground state distribution are also possible [40]. Using the methodology of section 4 here, we found -1.661 a.u. for the energy of the TDQMC ground state, to be compared with the result of -1.665 from the “exact” Ehrenfest molecular dynamics, and -1.657 from the Hartree-Fock calculation. The TDQMC average distance between the nuclei at ground state is 1.645 a.u. while it is 1.650 a.u. for the "exact" Ehrenfest molecular dynamics, and 1.636 for the Hartree-Fock approximation. Since in these calculations the system of electronic and nuclear walkers and guiding waves relaxes self-consistently to equilibrium we do not plot here the dependence of the ground state energy on the internuclear distance.

Next, we compare the results from the TDQMC molecular dynamics with the "exact" Ehrenfest MD and with the time dependent Hartree-Fock (TDHF) approximation, for the time-dependent ionization of hydrogen molecule in an external laser pulse. Dipole approximation can be used for non-relativistic motion of the particles and for wavelengths larger than the size of the molecule, where $V_{ext}(\mathbf{r}_i,t) = -e\mathbf{r}_i \bullet \mathbf{E}(t)$ and $V_{ext}(\mathbf{R}_I,t) = e\mathbf{R}_I \bullet \mathbf{E}(t)$ in Eq. (14), Eq. (15) and Eq. (25). Here $\mathbf{E}(t) = \mathbf{E}_0(t)\cos(\omega t)$ is the electric vector of the external optical field. In these calculations we use linearly polarized electromagnetic pulse at wavelength 335 nm with peak intensity 9 $10^{14}$ W/cm$^2$, whose time profile is shown in Fig. 1. In Fig.2 the result for the time-dependent ionization of $H_2$ obtained from the TDQMC molecular dynamics calculation are presented, compared with the "exact" Ehrenfest MD, and with TDHF approximation. The cases of quantum-classical and quantum-quantum TDQMC calculations are depicted. The ionization is assessed to be 1-$P_s$ where $P_s$ is the "survival" probability that the walkers for both electrons remain at a distance smaller than 10 atomic units from the nuclei. It is seen from Fig. 2 that the final ionization is slightly higher (within 5%) for the TDQMC calculations compared to the exact Ehrenfest, while the TDHF ionization is lower by about 20%. Clearly, the higher ionization in the TDQMC model is due to enhanced electron-electron and electron-nuclear correlations. Figure 3 shows that ensemble-averaged internuclear distance for the above cases. It is seen that the nuclei move apart slower in the "exact" Ehrenfest molecular dynamics than in TDQMC, which can be explained to be due to the lack of diffusion in the nuclear ensemble in the former. The underestimated electron-electron correlations in TDHF causes the molecule to be vibrationally excited but not to undergo dissociation, as seen in Fig.3. The TDQMC ground state distributions in 2D configuration space for the nuclei and for the electrons are depicted in Fig. 4 (a) and Fig. 4 (b), respectively. Figure 4 (c) shows the nuclear ensemble after the optical pulse, where different groups of the walkers correspond to neutral, ionized and dissociated molecules. There are double-ionized (bare) nuclei which move faster than the single-ionized hydrogen molecules and from the neutral hydrogen atoms which survive after the pulse (see Fig. 4 (d)). This is possible in the non-adiabatic models we use here where additional electrons are stripped away at certain instants of the optical field, mostly due to above threshold ionization close to the peak of the pulse, which adds some extra energy to the fragments. The non-local correlation length for the electrons increases on average from 2 a.u. for the ground state up to about 5 a.u. after the optical pulse, which is attributed to the spatial spread of the ionizing electronic walkers. It is seen from Figures 2 through 4 that although the ground state energy for a molecule with correlated electrons and nuclei is close to the Hartree-Fock result, the ionization and dissociation dynamics differ significantly due to the different dynamic correlations in the two cases. We also studied the orientation dynamics of two-dimensional hydrogen molecule under the influence of optical pulse with duration 15 fs and peak intensity 2 $10^{14}$ W/cm$^2$ at wavelength 800 nm, whose polarization is oriented at 45 degrees with respect to the initial axis of the molecule. Figure 5(a) shows the walker distribution for the electrons and the nuclei in 2D physical space in quantum-classical approximation where at the beginning the two protons are along the x-axis. Figure 5(b) shows the evolved walkers after the pulse where it is seen that the molecular axis has been re-oriented along the polarization of the external field while the molecule became almost dissociated.

**6. Conclusions**

In this paper, time dependent quantum Monte Carlo methodology is developed for quantum system of electrons and nuclei interacting through potentials. The quantum dynamics is modeled using ensembles of walkers and guiding waves which obey coupled linear time dependent Schrodinger equations for both the electrons and nuclei or, alternatively, by using regular bundles of classical trajectories for the nuclear degree. The quantum non-locality is incorporated into the model through effective potentials which are efficiently calculated by Monte Carlo integration. Unlike other many-body quantum methods TDQMC does not involve calculation of overlap, exchange and correlation integrals, which significantly improves its scaling properties. It also uses explicit Coulomb potentials instead of parameterized exchange-correlation potentials. The calculation of quantum potentials, which has been a major bottleneck for all particle methods, is avoided in TDQMC. The model calculations for the dynamics of low-dimensional hydrogen molecule in external field reveal that the quantum-quantum and quantum-classical predictions for the ionization and for the internuclear distance are slightly enhanced as compared with the results from the “exact” Ehrenfest molecular dynamics, but they differ significantly from the TDHF results. On the interpretational side, TDQMC differs significantly from the Bohmian mechanics in that the quantum potential plays a key role in the latter [41], while it is avoided in the former. The TDQMC method is easily parallelized and requires little communication between the processors only for calculation of the non-local quantum correlation effects.

**Acknowledgments**

The author gratefully acknowledges support from the National Science Fund of Bulgaria under contract WUF-02-05.

**Figure captions:**

**Figure 1**. Time dependence of the electric field used in the calculations, for carrier frequency $\omega=0.137$ a.u.

**Figure 2.** (Color online) Time dependent ionization for one-dimensional hydrogen molecule in an external optical pulse with carrier frequency $\omega=0.137$ a.u. (335 nm) and peak intensity $9\ 10^{14}$ W/cm$^2$. Black solid line- exact Ehrenfest molecular dynamics, gray (blue) line- quantum-classical TDQMC result, light-gray (red) line- quantum-quantum TDQMC result, dashed line- TDHF.

**Figure 3**. (Color online) Time dependent average internuclear distance for one-dimensional hydrogen molecule in an external optical pulse with carrier frequency $\omega=0.137$ a.u. (335 nm) and peak intensity $9\ 10^{14}$ W/cm$^2$. Black solid line -exact Ehrenfest molecular dynamics, gray (blue) line- quantum-classical TDQMC result, light-gray (red) line- quantum-quantum TDQMC result, dashed line- TDHF.

**Figure 4**. Dissociation of 1D hydrogen molecule. Distribution of the Monte-Carlo walkers in configuration space for the ground state: (a)- nuclei; (b)- electrons, and after the optical pulse: (c)- nuclei; (d)- electrons.

**Figure 5**. (Color online) Dissociation of rotating 2D hydrogen molecule. Distribution of the Monte-Carlo walkers in 2D physical space for the ground state (a) and after the optical pulse (b). The nuclei are represented by large light gray (red) dots while the electrons are represented by small gray (blue) dots.